\documentclass[graybox]{svmult}

\usepackage{type1cm}        
\usepackage{makeidx}         
\usepackage{graphicx}        
\usepackage{multicol}        
\usepackage[bottom]{footmisc}

\usepackage{newtxtext}       %
\usepackage[varvw]{newtxmath}     

\makeindex             

\def\cL{{\cal{L}}}

\def\cO{{\cal{O}}}

\def\be{\begin{equation}}
\def\ee{\end{equation}}
\def\beq{\begin{eqnarray}}
\def\eeq{\end{eqnarray}}
\def\bn{\begin{eqnarray*}}
\def\en{\end{eqnarray*}}

\def\s{\sigma}

\def\l{\lambda}

\def\g{\gamma}

\def\pd{\partial}
\def\e{\epsilon}

\def\m{\mu}

\def\r{\rho}

\begin{document}

\title*{Ultralight Dark Matter - A Novel proposal}
\author{T R Govindarajan} 
\institute{T R Govindarajan \at The Institute of Mathematical 
Sciences, Chennai, 600113 \\
Krea University, Sricity, 517646 \\ 
\email{trg@imsc.res.in,
govindarajan.thupil@krea.edu.in}}
%
%
\maketitle

\abstract{A novel proposal is made to account for the dark matter
component of the Universe. Ultralight dark matter with mass 
$\leq {\cal{O}}(10^{-22})~eV$ is one of the strong candidates for the 
missing mass which aids the formation of galaxies as well as holding 
them together. They are also known as fuzzy dark matter(FDM) which 
will come under Cold Dark matter. The question is what is this 
particle and its implications. How do we experimentally see it 
is an outstanding question. We propose to answer some of these
questions with some evidences and the estimates.}

\section{Introduction}
\label{sec:1}
For the past five decades the origin and the question of dark matter
\cite{dm} which constitutes 
25\% of matter in the universe have puzzled 
both astrophysicists, particle physicists  together. The 
observation that the radial 
speeds of luminous stars and other objects do not go down 
sufficently fast over a large distance in galaxies as expected  
through Newtonian/Einsteinian  gravity  ignited 
physicists to conjecture that there could be matter within the 
galaxies with only 
gravitational interactions. In addition galaxy clusters, 
gravitational lensing, bullet clusters and 
CMB spectrum \cite{dm}offer 
further support to this conjecture. There were several 
candidate proposals inspired by Beyond Standard Model (BSM) expectations.
Supersymmetry provides several potential massive particles. 
Sterile neutrinos and primordial blackholes too were considered
as potential candidates. There is also the suggestion of an extra 
dark photon may be lurking without any interaction with matter 
in the Universe except through gravity.
Unfortunately none of them satisfactorily answer all the queries 
regarding this. 

The desperation even led to the conjecture that gravity itself will 
need changes and Newtonian law of Universal gravitation might 
fail when the acceleration is below a 
particular threshold \cite{mond}.
This theory known Modified theory of Newtonian dynamics is also 
contender which explains rotational curves. But it fails in 
explaining bullet clusters or gravitational lensing in addition 
to being arbitrary.

Extremely light bosons from QCD axions\cite{axion}, 
and axion like scalar 
bosons have also been considered. While QCD axions have not 
succeeded in providing formulation as well as discovery such 
cold dark matter (CDM) can be potential candidate with ultra light 
mass. 

Having given the background to the dark matter question we will 
recollect one of the profound questions raised by Schrodinger 
in 50's namely `Must the photon mass be zero?'\cite{schrodinger}. 
We will 
deal with his answer and further developements on this question.
Finally we will link the dark matter question with this and 
provide a novel proposal for this burning question. 

\section{Schrodinger and Photon mass}
\label{sec:2}
Erwin Schrodinger in 1955 at Dublin Institute of Advanced Studies
posed the question about the photon mass. Since the massive photon 
will have three degrees of freedom whereas massless one will have 
one less, will it affect for example the Stefan's constant value.

While calculating in blackbody radiation,
energy density as a function of frequencies we
multiply by a factor of 2 to account
for the transverse degrees of freedom. Density of modes is given by:
\be
\r(\nu)~=~\frac{2h\nu^3}{c^2}\frac{1}{e^{h\nu\over kT}-1}
\ee
We can obtain Stefan's constant by integrating the density 
$$E~=~\int \r d\nu~=~ \s T^4$$


Schrodinger himself answered if the vector potential for the photon is coupled
only to a conserved current it will have very little effect if the mass of the photon is very tiny..
That is if the interaction is 
\begin{equation}
H_{int}~=~\int d^4x~j^\mu~A_\mu~,~~and~~\partial_\mu~j^\mu~=~0
\end{equation}
then corrections to the cross sections due to the longitudinal 
photons will be extremely small if the mass is very small..

\begin{svgraybox}
\large{Schrodinger: {\em Even if we find in Nature the 
limiting case is realized, we should still
feel the urge to adumbrate a theory which agrees 
with experience on approaching to the limit, not by a sudden}.
}
\end{svgraybox}
 
He went ahead and estimated the mass of the photon from the data 
available at that time about geomagnetic fields on the surface 
of earth. The limit he obtained was:
\begin{equation}
m_\gamma~~\leq~~10^{-16}eV
\end{equation}

The scale of this mass is easily understood as the only length scale 
available was the radius of the earth and we can expect 
the mass in natural units to be the inverse of radius. 
More modern data about geomagnetic fields satellite based 
experiments lead to an improvement to the value as 
$m_\gamma~\leq ~10^{-18}eV$.  Particle data book provides this as 
the current upper bound \cite{PDG}
\paragraph{Proca and Stueckelberg theory}
Massive vector boson theory known as Proca theory has two problems
(i) It has additional degree of freedom and massless limit has a
disconituity in the number of degrees of freedom. (ii) Mass term explicitely breaks local gauge 
invarince. But the massive QED is renormalisable and the extra
contributions are small if the mass is small\cite{banks}. But 
gauge invariance is our guiding principle and we expect it to be 
preserved. Stueckelberg theory avoids simultaneously 
the discontinuity in the degrees of freedom and  
the lack of gauge inavariance nicely\cite{stueckelberg}. 

Stueckelberg introduced another scalar field  $\phi$ and obtained a 
gauge invariant massive QED whose 
massless limit in addition avoids the 
discontinuity. The Lagrangian for Stueckelberg theory is:
\be \cL~=~-\frac{1}{4} \left(F_{\mu\nu}\right)^2~+~
\frac{1}{2}m^2\left(
A_\mu~-~\frac{1}{m}\pd_\mu\phi\right)^2~+~\bar\psi\Bigl[\g^\mu(i\pd_\mu
~+~eA_\mu)~-~M\Bigr]\psi
\ee
where $\phi$ is the Stueckelberg field and $\psi$ is the electron 
field. The gauge transformations are:
\be
\psi~\rightarrow e^{i\l(x)}\psi,~~A_\mu~\rightarrow A_\mu~-\pd_\mu
\l(x),~~\phi~\rightarrow \phi~+~m\l(x)
\ee
We can fix The gauge using: $-~{1\over 2}(\pd_\mu A^\mu~+~m~\phi)^2$.
 
We can also provide mass to the photon 
by the well known Higgs mechanism 
by coupling the abelian gauge field to a complex scalar field $\Phi$.
This will have nonzero vacuum expectation value 
giving mass to the the photon. We can write in the symmetry 
broken phase $\Phi~=~R~e^{i\phi}$. And the Lagrangian becomes:
\be
\cL~=~-\frac{1}{4}~F\wedge F~+~|D_\m \Phi|^2~-~V(\Phi)~+~\cdots
\ee
Phase of this field will be the Stueckelberg field
and this mechanism in a specific limit of freezing
the fluctuations of $R$ (i.e.,make $R$ very massive)
gives the Stueckelberg theory. There are other mechanisms like 
topological mass term by coupling to  Kalb Ramond field  $B_{\m\nu}$
which will not be discussed here. 
\section{Stueckelberg field as the dark matter candidate}
We learn the lesson that, Stueckelberg field does not interact with normal matter in a 
Compton length scale of $\frac{h}{mc}$. But it has energy and 
contributes extra terms to the energy momentum tensor. This 
can take part in gravitational interaction. The question 
we would like to ask is:  can it contribute as dark matter? 

If the mass of the candidate is too small, 
they will travel very close to the velocity of `light'
and decouple very early after the bigbang during radiation
dominated era. That will not help.That is where Bose and  
Einstein come to our rescue. Bose \cite{bose} wrote his famous paper in 1924 
providing the basis for his statistics which was followed 
by Einstein for a new phase of matter known as the `Bose Einstein 
Condensate (BEC)'\cite{bec}.

In our proposal  we may treat the
constituent particles as the Stueckelberg particles
which do not interact with matter. This can also be 
taken as the longitudinal 
photons through a gauge choice. These particles
will be such a candidate
only if they form a Bose-Einstein condensate.
The formation of a Bose-Einstein condensate
needs (i) a conservation law for particle number
(ii) the system should be at a temperature below the
critical temperature $T_c$ \cite{critical}.
For massless scalar fields, the conserved quantity is by
shift invariance of the field, which is broken by a mass term
and self interactions. It is an approximate symmetry.
The $T_c$ is given by
(where $\rho$ is the number density of a gas of particles),
\begin{equation}\label{eq2}
T_c =  \frac{\hbar c}{k_B}
\left(\frac{\rho\pi^2}{m_{\gamma}\zeta(3)}\right)^{1/3}.
\end{equation}
Ideal Bose gas is a quantum state of matter similar to ideal
gas in classical statistical mechanics. They obey Bose statistics 
and have integer spin. This was originally proposed by Bose for 
photon gas and extended to massive particles by Einstein.

While at large temperature they behave similarly, but at very
low temperatures they form a condensate 
under certain conditions due to
Bose Statistics. (After seeing Bose's letter, Einstein realised
immediately such a state can exist!)
BEC was obtained for Rubidium 87 gas whose mass is 86 amu
($\approx \cO(86)~Gev$).
Critical temperature is $\approx$ nano kelvins. We have very low 
value, because of high mass of the atoms, making up the gas.

Stueckelberg particles on the other hand 
have mass $\leq  10^{-20}eV$.
This makes the critical temperature very high
(close to the temperature of post Big bang after inflation)
\subsection{Dark matter and BEC}
We need to provide the relic density and the change in
density effected as a consequence of the Friedman expansion.
If dark matter is given by an initial density
of $\rho_0$ at the time of decoupling, before the radiation
dominated era begins we will know the present density. 
The epochs are given in table (\ref{tab}):

\begin{table}\label{tab}\caption{Epochs}
\begin{svgraybox}
\begin{tabular}{||p{5cm}|p{2cm}|p{3.6cm}||}\hline
~~Epoch and Time &~~Scale Factor &~~~~Temp.\\ \hline
\vfill
~~Radiation Era $1s$ to $1.2\times 10^{12}s$ &
\vfill~~~$\propto t^{1/2} $ & 
\vfill~~~~$10^{12}K$-$10^{4}K$\\ \hline
\vfill
~~Matter Era $4.7 \times 10^{4}y$ to $9.8 \times 10^{9}y$ & 
\vfill~~~$\propto t^{2/3}$ &
\vfill~~~~$10^{4}K$ - $4K$ \\ \hline
\vfill
~~Dark Energy $9.8 - 13.8$ billion y &
\vfill~~~$\propto e^{Ht}$ &
\vfill~~~~$~<~4K$ \\ 
\hline
\end{tabular}
\end{svgraybox}
\end{table}

Using these, we can relate the current density $\rho_{final}$ 
of dark matter to the initial density as:
\begin{equation}\label{eq3}
\rho^{\frac{1}{3}}_{final} = \rho^{\frac{1}{3}}_{0}
\frac{1}{(1.2\times 10^{12})^{\frac{1}{2}}}
\left(\frac{47000}{9.8\times 10^{9}}\right)^{\frac{2}{3}}
\frac{1}{1.377}.
\end{equation}
Employing this we get the relation between the observed
dark matter density $\rho$ and the critical temperature required
to achieve Bose-Einstein condensation ($m_\g$ in eV), 
\begin{equation}\label{eq4}
\rho~\sim~10^{-22}m_\gamma~T_c^3
\end{equation}

In SI units, the observed dark matter density in our galaxy 
\cite{PDG} which is approximately $\sim 10^{-22} kg/m^3$ or 1 proton/cc is 
recovered if we take
$m_\gamma \sim 10^{-19}eV$ and $T_c \sim 10^{17} K$.
The corresponding estimate for $m_\gamma~=~10^{-22}$ eV would
be $T_c~\sim 10^{19}K$.

Further the condensate once it is formed at the earlier times
will remain so for all epochs since the temperature has been 
cooling all the time.

For our galaxy, Milkyway luminous matter is $\approx~10^{10}$
solar masses. But dark matter is estimated to be $\approx~10^{12}$
solar masses.Dark matter (DM) halo is a theoretical 
model of galaxy that bounds the
galactic disc and extends beyond luminous (visible) part.
The mass of the halo dominates the galactic mass (nearly 95\% !) and
only postulated through observation of rotation curves.
DM halos are crucial for galaxy formation and evolution.

During galaxy formation temperature of matter is too high to
form gravitationally bound objects. Prior formation DM is needed
to add additional structure.
After the galaxy formation it extends far beyond observable
part and required for understanding velocities and lensing.

\subsection{DM Halo}
We have assumed that the current dark matter density can be
entirely explained in terms of Stueckelberg particles.
Several density profiles are in the literature.
There are two crucial parameters for modelling dark matter profile.
They are half radius and central density.
We can compute half radii of condensates and their masses.
With extremely ultralight mass for the particles,
we can consider the dark matter as a fluid.

A BEC condensate is there if the temperature is less
than $T_c$. This description is by the ansatz
$\phi \sim e^{-imc^2 t}\psi$ in a perturbed FRLW universe.
\begin{equation}
i\left(\partial_t + \frac{3}{2}\frac{\dot{a}}{a}\right)\psi =
\left(\frac{-\nabla^2}{2m} + mV\right)\psi
\end{equation}
where $V$ is a gravitational potential, in the linear approximation.

Following Hui, Ostriker,Tremaine, Witten \cite{witten}
the expression (Eqs.29,30 in the paper) for the 
half radius and mass are given by:
\begin{equation}\label{eq 6}
r_{\frac{1}{2}} = 3.925 \frac{\hbar^2}{GMm_{\gamma}^2},
\end{equation}

\begin{equation}
\rho_c = 4.4\times 10^{-3} \left(\frac{Gm_{\gamma}^2}{\hbar^2}\right)^3 M^4,
\end{equation}
where $\rho_c$ is the central density of the halo and $M$ is the
mass of the soliton.

We can parametrize mass as $10^{-17-x}eV$ and
$T_c~=~10^{\frac{50}{3}-y}K$ and obtain 
the sample space for these parameters.

There are couple of points we should remember about the program.
First if we consider the abelian Higgs model and obtain the 
Stueckelberg theory only in the limit, there will be further 
interactions of the Higgs field $R$ and the longitudinal 
component  $\phi$. Secondly due to the fluid/wavy behaviour the 
central density will avoid the well known core-cusp problem
which plagues in general all DM models particularly for the 
dwarf galaxy. 
\section{Arguments In favor}
Now we will see some arguments in favor of the proposal.
First there are several dwarf galaxies orbiting Milky way of 
various sizes from 200 light years to $10^5~$ light years. 
They hold stars from 1000 to $10^6$. 
They provide arguments in support of FDM.  Recently 
nano gravitational waves have been observed through radio 
telescopes arranged to analyse pulsar emission timings. This is 
known as Pulsar timing array (PTA). They surprisingly provide an 
argument in support of FDMs. Lastly one can check the Stress 
energy tensor due to massive vector theory. Due to the mass 
term there is extra contribution due to the longitudinal photons.
These are prominent when we consider long wavelength limit.
This provide another evidence in support of FDM. We will now elaborate each of these 
arguments. 
\paragraph{Dwarf galaxies} Stars are also at the 
initial stages of formation in some of the galaxies. 
N body Simulation with dark matter 
indicate the density grows exponentially at the core. But 
for FDM models there is no such problem. This is due to the fluid like 
behaviour of FDM. There are also 
anticpated around 200 such galaxies orbiting Milky way. But 
only about 40 have been seen so far. The rest are probably too faint
or DM content is too high and hence less luminous. There are several 
with diameters less than 1000 light years. The compton wavelength 
of the Stueckelberg particles will range from few light days to 
100-150 light years depending on the mass from $10^{-20}~-~10^{-24}
eV$. The size of the BEC condensate can be expected to be 
that of compton wavelength i.e., 
$\cO(\frac{h}{m_\g c})$. The size of smallest galaxy will 
set a limit on the 
compton wavlength and hence of the mass of the dark matter candidate. 
\paragraph{Segue-2 galaxy}
The smallest galaxy known so far is Segue-2.It is in the Aries 
constellation. It is $1.1\times 10^{5}$ light years away.It has a 
half radius of 115 light years. It has a mass of 
$\cO(10^6)M_{\odot}$. It holds around 1000 stars. 
We can expect 100 light years as the limit set by dwarf galaxy sizes
which will correspond to $m_\g \geq 10^{-24}eV$. This estimate 
also coincides with expectations from nano gravitational waves 
from pulsar timing array.
\paragraph{Pulsar timing array}
Pulsar timing array is essentially radio telescopes looking 
for Nano hertz gravitational waves due to merger of galaxies 
in the very early universe. This study is done by analysing the 
timing of the Pulsars within our galaxy.  
They have detected recently with confidence level {98\%} a
hum of the early universe.
This background noise is similar to CMB for Electromagntic spectrum.
This discovery has an interesting connection to FDM.
The gravitional waves of the galaxy mergers of the
early universe are affected by the dark matter between pulsars
in our galaxy and affect 
the noise. For this to be measured, the average distances
of Pulsars should be approximately the Compton wavelength of FDM
candidates. If this is correct, we expect the mass of FDM 
candidate should be $\geq 10^{-24}eV$ \cite{rubakov}. 
Recently with Parkes Pulsar Timing array which has been monitoring 
20 millisecond pulsars data for the period 2004-16 has been analysed.
it is claimed to produce a limit FDM mass should be $\geq 10^{-24}$.
Hence the window is narrowing for the FDM candidate proposals.
\paragraph{Massive photon and stress energy tensor}
Ryutov {\em et al} showed the Maxwell Proca electrodynmaics 
with finite photon mass changes stress tensor as expected. This is obviously
due to the extra longitudinal photons. This stress under certain
conditions develope a `negative pressure'\cite{ryutov}.
This negative pressure imitates gravitatational pull
and can become dominant.
The effect is associated with random
magnetic fields with correlation lengths exceeding
the photon Compton wavelength. The
stresses act predominantly on the interstellar gas
and cause an additional force pulling the
gas towards the center and towards the galactic plane.
This is precisely the scale at which the longitudinal
photons make the difference. But the force 
is not sufficient to explain the contribution to a star like sun
in the rotational curves. This might be lacuna of 
the Proca theory, but it requires to study this theory 
in the context of spontaneous symmetry breaking through 
Higgs mechanism and associate the extra preesure 
directly to the contribution of the Stueckelberg photons. 
\section{CMB and Stueckelberg field}
We started with photon mass query raised by Schrodinger and developed 
further to link up with fuzzy dark matter. Now we can go back and ask what 
can we learn from CMB radiation with photon mass. CMB provides evidence 
for hot big bang ends when radiation decouples with black body radiation 
at a temperature 2.7 K. While it agrees with black body curve to a great accuraacy 
but have shown discrepancies away from homogeniety. We can ask if we look for 
massive photon correction to the blackbody. Such a correction was worked out 
by Julien Heeck \cite{julien} and the energy density equation gets modified.

First we go back to the issue of the effect of
photon mass on black body radiation. Cosmic microwave background
being one of most perfectly known  `black body' we can check its effect
on the CMB distribution.
\begin{center}
\begin{figure}[h!]
\includegraphics[height=2in,width=3.5in]{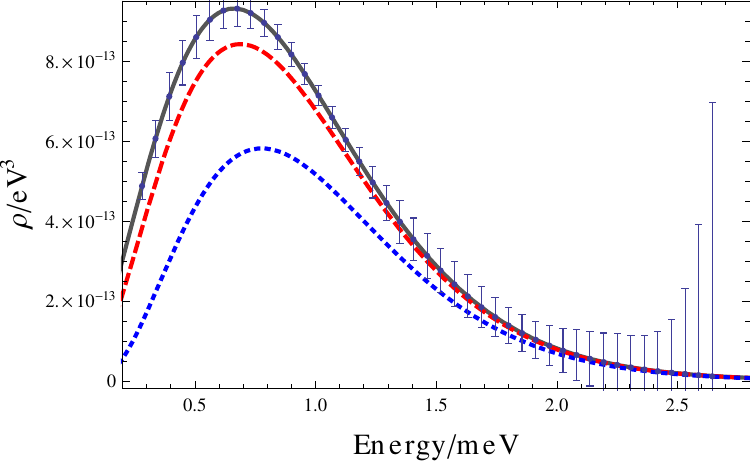}
\caption{Black body radiation with mass: Source: Phy Rev. Letts 111, 2013}
\label{cmb}
\end{figure}
\end{center}
One can put limits on the mass of the photon from
the deviation from the black body spectrum for CMB.
The density of photons with frequency is gievn by \cite{julien}:
\begin{equation}
\rho(E,T) ~ = ~\frac{E^3}{e^{E\over kT}-1}
\sqrt{1-\frac{m^2}{E^2}}
\end{equation} 
due to the modified dispersion relation $p^2~=~E^2~-~m^2$. See Fig.(\ref{cmb}).
Such an anlaysis was done and it gives poor estimate!
In the Fig.(\ref{cmb}) the gray line is for $m_\gamma~=~0$ and the 
red line is for $m_\gamma~=~10^{-5}$ eV. 
We expect $m_\g~<~10^{-6}eV$. Geo and Solar Magnetic field analysis give much
tighter bound as of now.

Obviously the limit is far removed from the limits we obtain from Earth and Solar magnetic 
field and the current expectation as candidate from FDM. This gives 
the realisation that the part of anisotropies can be atleast characterised 
by the mass paratmeter. But we have to consider Abelian Higgs model and the role 
of the interaction between Higgs and the Stueckelberg field also in the description.
This is ongong and will be reported.
 
\section{Detection of longitudinal photons}
Several attempts to estimate the mass of the photon have already 
been done. See \cite{PDG}. 
We can ask whether any particle physics signals can be provided 
for the longitudinal component of the photon. 
One such could be the implications 
for theorem due to Landau and Yang\cite{landau}. The theorem predicts 
a spin - 1 massive particle cannot decay to two massless spin -1 
particles. This implies the Z boson of Weinberg-Salam model 
cannot decay to two photons. Proof of the theorem does not require 
action or Hamiltonian. It simply follows from representation theory 
of Poincare group and Bose statistics of photons. It follows from 
the fact that symmetric product of two  massless representations of
Poincare group does not contain the massive
spin -1 representation \cite{bal}. 
This is easy to see in the rest frame of massive spin -1 particle 
we have two polarisations of photons $\e_i,i=1,2$  and $k,-k$ 
the momenta of outgoing photons along with spin vector $\vec{S}$ 
of the massive particle. We cannot write a non zero  amplitude
which is under $ k \leftrightarrow -k, \e_1 \leftrightarrow \e_2$.
(we need $k\cdot \e_i~=~0$ also). If the photon has mass then there
will also be longitudinal component, making it possible to violate 
the theorem. In actual process involving one of the photons with
longitudinal component, we will have non zero amplitude, but 
detection of the that photon will be very difficult. We have to look
for missing energy and momentum processes.
\subsection{Extension to WS Model with massive photons}
We can easiliy extend the above scheme to the standard model 
of Weinberg Salam $SU(2)_L\otimes U(1)$ theory. The action can
be written as 

\begin{equation}
\mathcal{L} = \mathcal{L}_g + \mathcal{L}_f + \mathcal{L}_S,
\end{equation}

\begin{equation}
\mathcal{L}_g  = -\frac{1}{4}B_{\mu\nu}B^{\mu\nu} - 
\frac{1}{4}Tr(f_{\mu\nu}f^{\mu\nu}) + \frac{1}{2}m_{\gamma}^2
\left(B_{\mu} - \frac{1}{m_\gamma}\partial_\mu \phi\right)^2,
\end{equation}
\begin{equation}
\mathcal{L}_S = |D_{\mu}\Phi|^2 -
\lambda\left(|\Phi|^2 - \frac{f^2}{2}\right)^2
\end{equation}
where $B_{\mu}$ is used to denote the weak hypercharge
field and $D_\mu$ is the covariant derivative acting on the Higgs.
$B_{\mu\nu}$ is the hypercharge field strength and $f_{\mu\nu}$ is
the $SU(2)$ field strength.
     
\section{Summary and Conclusions}
Stueckelberg theory is a powerful precursor to Higgs model
and introduces mass to the photon without breaking gauge invariance.
The theory is renormalisable in the ultraviolet and does not 
have infrared problem due to finite mass. Interestingly it achieves
freedom from infrared question even in the limit of 
$m_\g \longrightarrow 0$. This is because it naturally maps to 
Faddeev Kulish theory of asymptotic coherent states in the massless
limit. This can be understood from the difference between little
group of massless and massive representations of Poincare group.
These are E(2) and SO(3). There is a natural Inonu Wigner contraction
through a parameter which is in our case the mass itself. 
The review by Henri Ruegg, Marti Ruiz-Altaba, \cite{ruegg}
"The Stueckelberg Field", provides all the necessary background.
The details of Stueckelbrg particle as dark matter candidate 
are provided in our contribution \cite{trg}.
Dvali et al.,  \cite{dvali} propose 
holography can be formulated in terms
of the information capacity of Stueckelberg degrees of freedom.
These degrees of freedom act as qubits to encode quantum
information. The capacity is controlled by the inverse 
Stueckelberg energy gap to the size of the system.
They relate the scaling of the gap of the boundary Stueckelberg
edge modes to Bogoliubov modes. Unfortunately our 
program cannot be easily extended to QCD as there is no 
Stueckelberg program for non abelian gauge theory which is unitary 
and renormalisable. Can we get massive gravity by similar analysis?
Answer seems to be `no'. Again unitarity queries seems 
to be in conflict. 

We want to close this summary with profound remarks due to Bernhard
Riemann on the ultimate structure of space time in his 
Habilitation talk \cite{riemann}. 
\begin{svgraybox}
\begin{center}{\bf
AXIOMS UNDERLYING THE BASIS OF GEOMETRY \\
MAY NEED CHANGES\\
AT INFINETESIMAL AND INFINITE LEVEL.\\
Bernhard Riemann}
\end{center}
\end{svgraybox}
\begin{acknowledgement}

I thank the organisers of the International Symposium 
on Recent Developments in Relativistic Astrophysics 2023.
Special thanks are due to Shubhrangshu Ghosh for a wonderful meeting.
I also thank the participants for discussions. Special mention 
about Baniprata Mukhopadhyay for enlightening me on issues on 
Astrophysics and Cosmology. SRM University at Gangtok 
was wonderful place for discussions.
Finally this paper\cite{zeldovich} has
some implications for the current work which will be followed later.

\end{acknowledgement}

\end{document}